\documentclass[11pt,a4paper]{article}
\usepackage[utf8]{inputenc}

\usepackage{microtype}
\usepackage{amsmath,amssymb,bm}
\usepackage{graphicx,booktabs,array,multirow}
\usepackage[top=2.5cm,bottom=2.5cm,left=2.4cm,right=2.4cm]{geometry}
\usepackage[labelfont=bf,font=small]{caption}
\usepackage{cite}
\usepackage{setspace}
\usepackage{xcolor}
\usepackage[colorlinks=true,linkcolor=blue!50!black,citecolor=blue!50!black,urlcolor=blue!50!black]{hyperref}
\newcommand{\Sr}{S_r}
\newcommand{\Sp}{S_p}
\newcommand{\ST}{S_T}
\newcommand{\Rc}{R_C}
\newcommand{\au}{\,\mathrm{a.u.}}
\newcommand{\dd}{\mathrm{d}}

\title{External-field-modified entropic exchange in confined quantum systems:\\ static versus time-dependent electric fields}
\author{Vinod Prasad\\[2pt]
\small Department of Physics, Swami Shraddhanand College, University of Delhi, Delhi-110036, India}
\date{\small Preprint, October 2026}

\begin{document}
\maketitle

\begin{abstract}
\noindent
We review how external electric fields reshape the information content of spatially confined atoms, and we contrast the response to a uniform static field with that to a monochromatic, time-dependent (laser) field. The central quantities are the Shannon entropies in position and momentum space, $\Sr$ and $\Sp$, and their sum $\ST=\Sr+\Sp$, which obeys the Beckner--Bia{\l}ynicki-Birula--Mycielski (BBM) bound $\ST\ge 3(1+\ln\pi)\simeq 6.434$. Using the confined hydrogen atom (CHA) and the particle in a spherical box (PISB), its tight-confinement limit, as reference systems, we first show that the field-free total entropy of the CHA depends non-monotonically on the confining radius whereas that of the PISB is strictly scale invariant. We then present Stark maps of $\Sr$, $\Sp$ and $\ST$ for the ten lowest $m=0$ levels at $\Rc=6\au$. The ground-state polarizability is reduced from the free-atom value $4.5\au$ to $4.06\au$ in the CHA but is $46.8\au$ in the PISB, and the entropic response of the two systems differs correspondingly: $\Sr$ of the CHA ground state \emph{increases} with the field while that of the PISB \emph{decreases}, with the two compensating contributions to $\ST$ nearly cancelling in both cases. Accidental degeneracies are lifted discontinuously by an infinitesimal field and produce an abrupt entropic exchange, and avoided crossings between levels of strongly different angular momentum give narrow gaps (down to $6\times10^{-4}\au$ for $\Delta l=5$) and nearly complete exchange of $\ST$. For the time-dependent case we use the non-perturbative quasi-energy (Floquet) formalism and define the entropies from the cycle-averaged density of the dressed state. At the one-photon $1s$--$2p$ resonance the Floquet branches undergo an avoided crossing whose minimum gap equals the Rabi splitting $E_0 d_{12}$ with $d_{12}=0.92\au$; across it the two branches exchange practically all of their $\Sr$ and $\Sp$ over a frequency window of about $2.7$ gap widths, and the total entropy of each branch overshoots the values of both parent states, showing that $\ST$ is non-additive under coherent superposition. Raising the laser amplitude from $0.01\au$ to $0.04\au$ widens and shifts the crossing (AC Stark effect) and broadens the entropic exchange. These results provide a common framework for comparing static and time-dependent field control of information measures in confined quantum systems.
\end{abstract}

\noindent\textbf{Keywords:} quantum confinement; Shannon entropy; avoided crossings; Stark effect; Floquet quasi-energies; AC Stark effect; entropic uncertainty relation

\tableofcontents
\newpage
\section{Introduction}
\label{sec:intro}

\subsection{Information measures in physics}
Since the foundational work of Shannon~\cite{b1}, information theory has found application well beyond communication engineering. It is now used in classical and quantum mechanics~\cite{b2,b3}, in statistical mechanics~\cite{b4,b5}, in quantum computation~\cite{b6,b7} and quantum information~\cite{b8,b9,b10,b11}, in quantum biology~\cite{b10}, in molecular biology and ecology~\cite{b13,b14}, in text and biomedical information retrieval~\cite{b12}, in signal processing, data compression and digital forensics~\cite{b15,b16,b17}, in nuclear collision dynamics~\cite{b11} and in the characterisation of directional emission from microcavity lasers~\cite{b22}. A large part of this activity concerns the quantification of uncertainty~\cite{b18,b19,b20,b21,b23}. In quantum mechanics the relevant information measures are most often entropic. The Shannon entropies in position space, $\Sr$, and in momentum space, $\Sp$, quantify how localized or delocalized a state is: a large entropy signals a broad probability density and hence a large uncertainty about the location (or momentum) of the electron, while a small entropy signals a compact density.

The sum $\ST=\Sr+\Sp$ is subject to the Beckner--Bia{\l}ynicki-Birula--Mycielski (BBM) inequality~\cite{b24},
\begin{equation}
 \ST=\Sr+\Sp\;\ge\; D\,(1+\ln\pi),
 \label{eq:bbm}
\end{equation}
where $D$ is the dimensionality of the space ($D=3$ throughout this work, so that the bound is $6.434$ in atomic units). Because it constrains the \emph{joint} localisation of the position and momentum densities, Eq.~(\ref{eq:bbm}) is regarded as a stronger statement of the Heisenberg uncertainty principle~\cite{b23}, and it has been examined for many quantum systems~\cite{b25,b26,b27}. In atoms, $\ST$ correlates with several other physical properties and tracks them consistently~\cite{b26}. A basic principle of information theory is that more knowledge about a system means lower entropy, whereas maximal ignorance, with all states equally probable, gives maximal entropy.

\subsection{Information, thermodynamics and the physical meaning of entropy}
The connection between information and thermodynamics has a long history. Brillouin~\cite{b28,b29} introduced negentropy, or negative entropy, as the quantity that links thermodynamic and informational entropy, and used it to resolve the paradox of Maxwell's demon~\cite{b28,b30,b31,b32,b33}. In that thought experiment a demon sorts fast from slow molecules and so creates a temperature difference that could drive a heat engine; the apparent reduction of entropy is paid for by the information the demon acquires. Szilard~\cite{b34,b35} later proposed a model in which the demon's information is converted into mechanical work, the Szilard engine~\cite{b36,b37}, which extracts at most the work equivalent to one bit of information. This is historically the first quantitative relation between energy and information.

Gabor~\cite{b38,b39,b40} adopted the same negentropy concept. In Brillouin's formulation negentropy corresponds to the \emph{grade} of energy in Kelvin's principle of degradation of energy: a system with temperature differences between its parts, or with pressure differences, contains negentropy that may be turned into work. Brillouin~\cite{b29} and Abe and Okuyama~\cite{b5} related the statistics of negentropy to thermodynamics through Carnot's principle, and Ref.~\cite{b41} gives a detailed account of the link between information theory and physical systems based on Landauer's principle. Landauer~\cite{b42} stressed that information is a physical quantity and related it to energy dissipation (heat loss) when known or unknown information is erased from a physical device~\cite{b43,b44}. His principle states that heat is dissipated to the environment only on erasure, with a minimum cost per bit that is tied to the Shannon limit for the minimum energy needed to produce one bit~\cite{b45}. Jaynes~\cite{b46} showed that the statistical entropy equals $k_B\ln 2$ times the maximum missing Shannon information of a closed system subject to a mean-energy constraint~\cite{b41}; each admissible state is then equally likely, which is the basis of the statistical entropy. Reviews of the many definitions of entropy and of their mutual relations are available in Refs.~\cite{b47,b48}.

\subsection{Entropic measures of atoms and confined systems}
Information theory has attracted interest in quantum systems over the last two to three decades. Cerf and Adami~\cite{b27} studied negative conditional entropy between entangled systems as an indicator of quantum non-separability. Guevara \emph{et al.}~\cite{b26} showed that the total Shannon entropy of neutral atoms is a correlation measure that depends on the ionization energy and the atomic number, and found a relation between the entropy sum and the correlation energy of a weakly inhomogeneous electron gas. Gonz\'alez-F\'erez and Dehesa~\cite{b49} pointed out that $\Sr$ and $\Sp$ are exchanged between the two states involved in an avoided crossing in parallel electric and magnetic fields, at the same field value at which the energy levels repel, so that the Shannon entropy is an indicator of avoided crossings. Saha and Jose~\cite{b50} used the Shannon entropy to identify correlation and relativistic effects in the endohedral atom Ba@C$_{60}$. Mart\'inez-S\'anchez \emph{et al.}~\cite{b51} computed the entropy of the hydrogen atom for four different confining potentials. Flores-Gallegos~\cite{b52,b53} showed that the entropy displays the periodicity and shell structure of the atoms, changes abruptly on passing from one shell to the next, and is almost independent of the method used to compute it. Est\'a\~n\'on \emph{et al.}~\cite{b54} found that information measures of the confined helium atom are sensitive to the correlation factor and may help to quantify electron correlation, and Nascimento \emph{et al.}~\cite{b55} showed that the coupling and decoupling of double quantum dots can be monitored with the Shannon entropy. Shannon entropy and related measures of confined systems have been studied intensively during the last decade~\cite{b56,b23,b57,b58,b54,b55,b59,b60,b61,b62,b63,b64,b65,b25,b66,b67,b68,b69,b70,b71,b72,b73,b74,b75,b76}.

The Shannon entropy is the most widely used measure, but not the only one. Others are the Onicescu information energy~\cite{b77,b78,b79,b80,b81,b82}, the Kullback--Leibler (KL) distance~\cite{b83,b84,b85,b86}, the R\'enyi entropy~\cite{b82,b80,b87,b88,b89,b90}, the Tsallis entropy~\cite{b80,b82,b88,b89,b91,b92,b93,b94,b95} and the L\'opez-Ruiz--Mancini--Calbet (LMC) complexity~\cite{b96,b97,b98,b99}. The Onicescu energy is large for a localized density and small for a delocalized one, so it behaves inversely to the Shannon entropy. Chatzisavvas \emph{et al.}~\cite{b83} used these measures to expose shell effects in atoms: the Onicescu energy has local maxima and $\Sr$ local minima at closed-shell atomic numbers $Z$, and $\ST$ is approximately linear in $\ln Z$. The KL distance (relative entropy) measures the difference between two probability distributions and vanishes only if they coincide. The R\'enyi entropy quantifies concentration within a distribution and distinguishes regions of low and high probability, and is also used as a measure of correlation. The Tsallis entropy measures the spread or concentration of the density. The LMC complexity combines concentration and spread, being the product of the Onicescu energy and the exponential of the Shannon entropy in the respective space.

\subsection{Spatial confinement}
The influence of confinement on quantum systems was first examined in 1937, when Michels \emph{et al.}~\cite{b100} studied the effect of high pressure on the polarizability of atoms and molecules, down to radii of $5\au$. Shortly afterwards Sommerfeld and Welker~\cite{b101} considered the $1s$ state of hydrogen in an impenetrable spherical cavity. Its energy rises as the confining radius shrinks and becomes positive below the critical radius $R_C^0=1.835\au$, at which the electron is bound only by the external wall and the Coulomb attraction is no longer decisive. Critical radii of excited states were later reported by Varshni~\cite{b102}; the experimental and theoretical situation of that period is summarized in Ref.~\cite{b103}. The effect of pressure on the physical and chemical properties of materials has been studied extensively~\cite{b68,b69,b70,b104,b105,b106,b107,b108,b109}. Confinement is also essential in astrophysics for plasma properties and dwarf stars~\cite{b110,b111,b112,b113} and for the interiors of Jupiter and Saturn~\cite{b114,b115}.

The hydrogen atom, being the simplest atom, has been studied under many types of confinement and by many methods~\cite{b116,b105,b106,b107,b109,b117,b118,b102,b76,b119}. Naturally occurring cavities, such as zeolite cages and fullerenes, and impurities in semiconductors, are common realisations of spatial confinement~\cite{b126,b127,b128,b129,b130}. Confinement can be used to tune energy levels, orbitals, polarizability, magnetic moments, photoionization and transition probabilities~\cite{b117,b120,b121,b122,b123,b56,b124,b125,b57}, through an external confining potential and/or through external static electric and magnetic fields and lasers~\cite{b118,b102,b126,b127,b128,b129,b130,b131,b132,b133,b134,b135,b136,b137,b138,b139}. Chakraborty and Chattaraj~\cite{b140} reviewed how confinement affects dynamics, reactivity and bonding. Burrows and Cohen~\cite{b104,b141} obtained the exact solution for the hard-wall CHA in terms of Kummer functions and studied polarizabilities and nuclear shielding factors; Aquino \emph{et al.}~\cite{b142} gave highly accurate energies and wave functions, Mukherjee \emph{et al.}~\cite{b60} treated hydrogen-like ions in plasma, Aquino~\cite{b143} used confined harmonic-oscillator functions to obtain energies, and Pang~\cite{b144} studied the hydrogen molecule by quantum Monte Carlo. Sen~\cite{b25} analysed the hydrogen atom in a spherical shell and the well-defined radial nodes of its states, and the $1s$ state of the CHA at $\Rc=2\au$ has the energy of the free $2s$ state although the wave functions differ~\cite{b145}. Further results on one-electron systems are collected in Ref.~\cite{b106}. The conditions for incidental and accidental degeneracies at radial nodes of the free functions were formulated by Pupyshev and co-workers~\cite{b146,b147,b148}: for instance the $2s$ and $3d$ levels, and the $3s$ and $4d$ levels, become degenerate at $\Rc=2\au$ and the $3p$ and $4f$ levels at $\Rc=6\au$. Baye and Sen~\cite{b149} used the Lagrange-mesh method for energies and polarizabilities of an H atom in a central potential. The three basic reference systems are the free hydrogen atom (FHA), the confined hydrogen atom (CHA) and the particle in a spherical box (PISB). Confinement potentials are either hard-wall (HW, infinite) or soft-wall (SW, finite).

\subsection{External fields}
External fields were among the first tools used to probe atoms. Zeeman's observation of the splitting of spectral lines in a magnetic field in 1896~\cite{b150}, and Stark's observation of the splitting of lines in a static electric field in 1913~\cite{b151}, were decisive for the development of quantum mechanics and each earned a Nobel prize. Electric and magnetic fields have since been applied to many systems~\cite{b152,b153} to examine their effects on energies~\cite{b154,b155,b156}, transition probabilities~\cite{b157,b121,b138,b134}, dipole moments~\cite{b158,b159}, oscillator strengths~\cite{b160,b161,b162} and polarizabilities~\cite{b163,b164,b165}. With the advent of lasers, light fields alone or combined with electric and/or magnetic fields became a further means of manipulation~\cite{b133,b166,b131,b132,b167}. Tsong~\cite{b168} described how an electric field can be used to manipulate and place atoms on a surface; Perelomov \emph{et al.}~\cite{b169} studied ionization of atoms in an alternating field and its dependence on frequency; Braun~\cite{b170} treated the splitting of Rydberg levels in combined electric and magnetic fields by semiclassical methods; Krems~\cite{b171} reviewed external-field control of chemical reactions, dissociation and collisions of ultracold molecules; and Delone and Krainov~\cite{b172} discussed the AC Stark shift of energy levels for resonant and non-resonant conditions within the Floquet quasi-energy picture. In a static electric field one finds Stark ladders and in a magnetic field Landau levels, and with both fields interesting combined effects. The effect of external fields on a particle in a box, quantum wells~\cite{b133} and quantum dots~\cite{b173,b174} has likewise been studied.

\subsection{Aim and organisation}
Information measures have been studied for free and confined systems, but their response to external fields has received much less attention. This question is relevant to quantum information science because the realisation of qubits depends on superpositions of states that are protected from the environment, and the entropic content of a field-dressed state is a measure of how it is delocalized. In this review we therefore compare, on equal footing, how a static electric field and a monochromatic laser field modify the entropic measures of confined atoms. Section~\ref{sec:measures} defines the measures and recalls the BBM bound. Section~\ref{sec:methods} summarizes the confined systems and the numerical machinery (boundary conditions, finite differences, matrix perturbation theory and Floquet theory). Section~\ref{sec:results} presents the results: field-free entropies (Sec.~\ref{sec:free}), static-field Stark maps (Sec.~\ref{sec:stark}) and laser-induced avoided crossings (Sec.~\ref{sec:laser}). Section~\ref{sec:compare} contrasts the two kinds of field and Sec.~\ref{sec:concl} gives conclusions and an outlook.
\section{Information-theoretic measures}
\label{sec:measures}
All quantities are in atomic units and logarithms are natural. For a normalized state $\psi(\mathbf r)$ with Fourier transform $\phi(\mathbf p)=(2\pi)^{-3/2}\!\int\psi(\mathbf r)e^{-i\mathbf p\cdot\mathbf r}\,\dd^3r$, the position and momentum densities are $\rho(\mathbf r)=|\psi(\mathbf r)|^2$ and $\gamma(\mathbf p)=|\phi(\mathbf p)|^2$. The Shannon entropies are
\begin{equation}
 \Sr=-\!\int\rho(\mathbf r)\ln\rho(\mathbf r)\,\dd^3r,\qquad
 \Sp=-\!\int\gamma(\mathbf p)\ln\gamma(\mathbf p)\,\dd^3p,\qquad
 \ST=\Sr+\Sp .
 \label{eq:shannon}
\end{equation}
A compact state has small $\Sr$ but, by the uncertainty principle, large $\Sp$; the sum is bounded from below by Eq.~(\ref{eq:bbm}). The bound is saturated by Gaussians, and for the lowest states of atoms $\ST$ lies close to it (Sec.~\ref{sec:free}).

Further measures are defined for a generic density $\rho$, and analogously in momentum space. The Onicescu information energy is $O=\int\rho^2\,\dd^3r$. The Kullback--Leibler distance of $\rho$ from a reference density $\rho'$ is $D_{KL}(\rho\|\rho')=\int\rho\ln(\rho/\rho')\,\dd^3r\ge0$, which vanishes only for $\rho=\rho'$. The R\'enyi and Tsallis entropies of order $\alpha$ are
\begin{equation}
 R_\alpha=\frac{1}{1-\alpha}\ln\!\int\rho^\alpha\,\dd^3r,\qquad
 T_\alpha=\frac{1}{\alpha-1}\Bigl(1-\int\rho^\alpha\,\dd^3r\Bigr),
\end{equation}
and both reduce to $\Sr$ for $\alpha\to1$. The LMC complexity is $C=O\,e^{\Sr}$. In this review the discussion of field effects is based on $\Sr$, $\Sp$ and $\ST$.

For the central potentials considered here and a field along $z$ the magnetic quantum number $m$ is conserved, and we restrict attention to $m=0$. A field-dressed state is then $\psi(r,\theta)=\sum_l R_l(r)Y_{l0}(\theta)$, and its momentum-space counterpart is $\phi(p,\theta_p)=\sum_l(-i)^l\tilde R_l(p)Y_{l0}(\theta_p)$ with $\tilde R_l(p)=\sqrt{2/\pi}\int_0^\infty r^2 j_l(pr)R_l(r)\,\dd r$, where $j_l$ is the spherical Bessel function. The entropies (\ref{eq:shannon}) were evaluated by quadrature of these expressions on radial, angular (Gauss--Legendre) and momentum grids.

\subsection*{Entropic exchange at an avoided crossing}
\label{sec:twolevel}
The physical content of the ``entropic exchange'' studied below can be understood from a two-level model. Let two diabatic states $|a\rangle,|b\rangle$ with energies $E_a(\lambda),E_b(\lambda)$ depend on a control parameter $\lambda$ (the field strength $F$ for the static case, the detuning for the laser case) and be coupled by a matrix element $V$. The adiabatic states are $|\pm\rangle=\cos\vartheta\,|a\rangle\pm\sin\vartheta\,|b\rangle$ (up to ordering) with
\begin{equation}
 \sin^2\vartheta=\tfrac12\Bigl(1-\frac{\delta}{\sqrt{\delta^2+\Delta^2}}\Bigr),\qquad
 \delta=E_b-E_a,\qquad \Delta=2|V| ,
 \label{eq:twolevel}
\end{equation}
where $\Delta$ is the minimum gap. Away from the crossing $|\pm\rangle$ coincide with the diabatic states and carry their entropies; at the crossing they are equal-weight superpositions. If $\delta$ varies linearly with slope $s$ near $\lambda_c$, the mixing angle changes between 10\% and 90\% over a window $\delta\lambda=8\Delta/(3s)\simeq2.67\Delta/s$. Because the entropies are not linear functionals of the density, the entropy of $|\pm\rangle$ is not a weighted average of $S_a$ and $S_b$: the cross term $2\sin\vartheta\cos\vartheta\,\psi_a\psi_b$ in the density can make the entropy of the superposition \emph{larger} than that of either constituent. We use this picture to interpret the numerical results below.
\section{Models and methods}
\label{sec:methods}

\subsection{Confined systems}
The Hamiltonian of an atom (or a particle) under spherical confinement is, in atomic units,
\begin{equation}
 H_0=-\tfrac12\nabla^2+V(r)+V_C(r),
 \label{eq:H0}
\end{equation}
where $V(r)=-Z/r$ is the Coulomb attraction ($Z=1$ for hydrogen), $V(r)=0$ for the PISB, and $V_C$ is the confining potential. For $V_C=V_C(r)$ the wave function separates, $\psi(r,\theta,\phi)=R(r)Y(\theta,\phi)$; the angular part is independent of the confining radius and only the radial part is affected by it. Confinement quantizes the spectrum and modifies energies and all derived properties relative to the free system. The only difference between the PISB and the CHA is the Coulomb term, so the PISB is the limit of very strong confinement in which the kinetic energy $\sim\Rc^{-2}$ dominates the Coulomb energy $\sim\Rc^{-1}$. Michels \emph{et al.}~\cite{b100} gave the first power-series solution of the confined atom. Two models are used: (i) the hard-wall CHA (infinite $V_C$ for $r>\Rc$) and (ii) the soft, attractive potential of a fullerene cage, e.g.\ H@C$_{36}$ or H@C$_{60}$, represented either by a square-well shell of width $\Delta$ or by a Gaussian potential of width $\sigma$~\cite{b119}. The quantitative results of Sec.~\ref{sec:results} are for the hard-wall CHA and the PISB.

\subsection{Boundary conditions}
The boundary conditions encode the physics of the confinement. \emph{Dirichlet} conditions (boundary conditions of the first kind, named after J.~P.~G.~Lejeune Dirichlet) prescribe the value of the solution on the boundary~\cite{b175,b176}. For a particle in a one-dimensional infinite well $0\le x\le L$,
\begin{equation}
 \psi(x)\big|_{x=0}=\psi(x)\big|_{x=L}=0,
\end{equation}
because infinite walls forbid the particle at the edges. For hard-wall spherical confinement correspondingly $\psi(r)|_{r=\Rc}=0$. \emph{Neumann} conditions (second kind, named after C.~G.~Neumann) prescribe the normal derivative~\cite{b177,b178},
\begin{equation}
 \left.\frac{\dd\psi}{\dd x}\right|_{x=0}=\left.\frac{\dd\psi}{\dd x}\right|_{x=L}=0,\qquad
 \left.\frac{\dd\psi}{\dd r}\right|_{r=\Rc}=0,
\end{equation}
and are used as the idealised model of a finite (soft) wall, where the particle has a non-zero probability outside the boundary. They lead to different spectra and wave functions than Dirichlet conditions and are more realistic for many problems. All results in Sec.~\ref{sec:results} use Dirichlet conditions.

\subsection{Finite-difference method}
The radial equation is discretized by finite differences~\cite{b179,b180,b181}. Weights are obtained from truncated Taylor expansions, and the leading truncation term gives the error order. With $2N+1$ stencil points the error is $O(h^{2N})$: the 3-point central formula is second order, $O(h^2)$, and the 9-point formula is eighth order, $O(h^8)$. Xie \emph{et al.}~\cite{b180}, for example, used a fourth-order scheme for telegraph equations. We use the 9-point central stencil, which reaches a given accuracy with far fewer grid points and hence at lower cost. Diagonalizing the resulting matrix gives the energies and the radial wave functions of each $l$ block, i.e.\ the field-free (bare) states, from which the matrix elements of the field operators are computed.

\subsection{Static electric field: matrix perturbation theory}
Many systems differ only slightly from an exactly solvable one, and the Hamiltonian is then written as $H=H_0+H_1$, where $H_1$ is the perturbation from an external field or an internal interaction. For a uniform static field $F$ along $z$ (length gauge), $H_1=Fz$. We expand the field-dressed states in the field-free eigenstates of the same system and diagonalize $H_0+Fz$ in this basis. This is the matrix form of perturbation theory: it is exact within the chosen basis and not restricted to small $F$. For $m=0$ the dipole operator couples only $l\to l\pm1$,
\begin{equation}
 \langle n\,l\,0|z|n'\,l{+}1\,0\rangle=\sqrt{\frac{(l+1)^2}{(2l+1)(2l+3)}}\int_0^\infty r\,u_{nl}(r)\,u_{n',l+1}(r)\,\dd r ,
 \label{eq:dipole}
\end{equation}
with $u=rR$. Without a field the parity of the state is conserved; a field mixes states of opposite parity and lifts the degeneracy in $m$, and the smaller the perturbation the sharper an avoided crossing between two levels, while larger perturbations round it off. This is the sense in which a field can be used to modify and tailor the states.

\subsection{Time-dependent field: Floquet theory}
For a monochromatic field of angular frequency $\omega$ the Hamiltonian is periodic in time and the Schr\"odinger equation is solved by the Floquet method~\cite{b182}. We consider a linearly polarized field along $z$, $H(t)=H_0+E_0\,z\cos\omega t$, treated as acting on the atom without back-action on the field, which is appropriate for a continuous weak field or a long pulse for which the interaction is adiabatic. Floquet states have the form $\Psi_\alpha(\mathbf r,t)=e^{-i\varepsilon_\alpha t}\sum_m\phi^{(\alpha)}_m(\mathbf r)e^{-im\omega t}$, where $\varepsilon_\alpha$ is the quasi-energy and $m$ the photon (harmonic) index. The harmonics satisfy
\begin{equation}
 (H_0+m\omega)\,\phi_m+\tfrac12E_0\,z\,(\phi_{m-1}+\phi_{m+1})=\varepsilon\,\phi_m ,
 \label{eq:floquet}
\end{equation}
which is diagonalized in the product basis of field-free states and harmonics $|n\,l\rangle\otimes|m\rangle$. The method is non-perturbative in $E_0$ and contains the AC Stark shift, multiphoton processes and counter-rotating terms. The density of a dressed state oscillates in time; we use its \emph{cycle average}
\begin{equation}
 \bar\rho(\mathbf r)=\sum_m|\phi_m(\mathbf r)|^2,\qquad
 \bar\gamma(\mathbf p)=\sum_m|\tilde\phi_m(\mathbf p)|^2,
 \label{eq:cycleavg}
\end{equation}
with $\sum_m\langle\phi_m|\phi_m\rangle=1$, in Eq.~(\ref{eq:shannon}) to define $\Sr$, $\Sp$ and $\ST$ of the dressed state.

\subsection{Computational details of the new results}
All calculations reported in Sec.~\ref{sec:results} are for a confining radius $\Rc=6\au$ (hard wall, Dirichlet), unless stated otherwise (the $\Rc$ dependence is shown in Fig.~\ref{fig:free}). The field-dressed states for $m=0$ are obtained in the basis of field-free states with angular momenta $l=0,\dots,5$ ($s$ to $h$); the energies and entropies of the lowest levels are converged with respect to the number of radial states per $l$ (four or more radial states per $l$ suffice). We follow the ten lowest levels, ordered by energy at each field value. Static fields cover $0\le F\le0.2\au$ in steps of $10^{-3}\au$. Laser calculations cover $0.364\le\omega\le0.524\au$ around the one-photon resonance between the two lowest levels, for amplitudes $E_0=0.01$ and $0.04\au$, with enough photon harmonics for convergence of the quasi-energies ($|m|\le7$ is more than sufficient). Levels are labelled by the hydrogenic notation $nl$ with $n=n_r+l+1$, where $n_r$ is the number of radial nodes. Two caveats apply to the static-field results. First, because the angular basis stops at $l=5$, levels above about the fifth (above the third for the PISB at the largest fields) are not converged with respect to $l_{\max}$ at large $F$, and features that involve them reflect the model space; in the weak-field region ($F\lesssim0.05\au$) levels up to about the eighth are insensitive to the truncation. Second, at $F=0$ the CHA levels 5 and 6 ($3p$ and $4f$) are degenerate, and a diagonalization at exactly $F=0$ returns an arbitrary mixture of them.
\section{Results}
\label{sec:results}

\subsection{Field-free entropies: confined hydrogen atom versus particle in a spherical box}
\label{sec:free}
\begin{figure}[t]
\centering
\includegraphics[width=0.95\textwidth]{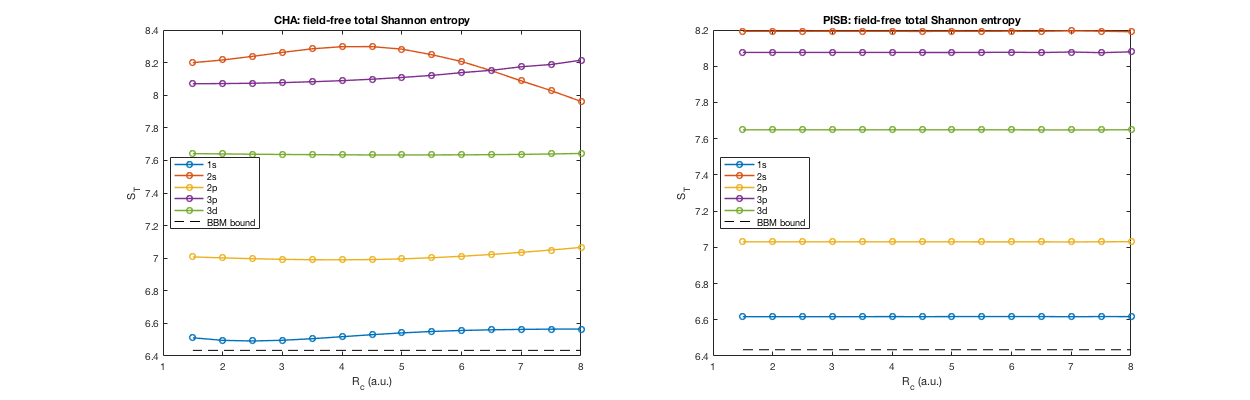}
\caption{Field-free total Shannon entropy $\ST$ versus confining radius $\Rc$ for (a) the hard-wall confined hydrogen atom (CHA) and (b) the particle in a spherical box (PISB). Dashed line: BBM bound $3(1+\ln\pi)=6.434$.}
\label{fig:free}
\end{figure}

Figure~\ref{fig:free} shows the total Shannon entropy of the $1s$, $2s$, $2p$, $3p$ and $3d$ states of the hard-wall CHA and of the PISB for $1.5\le\Rc\le8\au$ in the absence of a field. All values lie above the BBM bound $6.434$ (dashed line). The two systems behave in qualitatively different ways.

\emph{PISB.} The total entropy of every level is independent of $\Rc$ (the variation over the whole range is at most $5\times10^{-3}$, i.e.\ numerical noise). This follows from scaling: a box of radius $\Rc$ has wave functions $\psi(\mathbf r)=\Rc^{-3/2}f(\mathbf r/\Rc)$, so that $\Sr\to\Sr+3\ln\Rc$ while $\Sp\to\Sp-3\ln\Rc$, and the sum is invariant. The values $6.618\,(1s)$, $7.030\,(2p)$, $7.649\,(3d)$, $8.193\,(2s)$ and $8.076\,(3p)$ are therefore pure numbers characterizing the nodal structure; the ordering $1s<2p<3d<3p<2s$ reflects their nodal and angular structure.

\emph{CHA.} The Coulomb term breaks the scale invariance, so $\ST$ depends on $\Rc$, and does so non-monotonically. The $1s$ entropy has a minimum of $6.492$ near $\Rc\simeq2.5\au$, only $0.058$ above the BBM bound. For larger radii it increases, to $6.557$ at $\Rc=6\au$ and $6.566$ at $8\au$ on the way to the free-atom value, and for smaller radii it increases as well ($6.512$ at $1.5\au$) towards the PISB value $6.618$. The $2s$ entropy has a broad maximum of $8.298$ at $\Rc\simeq4$--$4.5\au$ and falls to $7.962$ at $8\au$ (a variation of $0.34$). The $2p$ entropy has a shallow minimum ($6.990$) at $\Rc\simeq4\au$, the $3p$ entropy increases monotonically from $8.070$ to $8.215$, and the $3d$ entropy is nearly constant ($7.633$--$7.643$), which we attribute to the nodeless $3d$ radial function being the least sensitive to the competition between the Coulomb attraction and the wall. For tight confinement the CHA entropies of the excited levels approach those of the PISB (differences of $0.02$ or less at $\Rc=1.5\au$ for $2s$, $2p$, $3p$, $3d$), consistent with the dominance of the kinetic energy; at $\Rc=8\au$ the differences reach $-0.23$ ($2s$) and $+0.14$ ($3p$), a direct measure of the Coulomb correction to the box entropy.

Table~\ref{tab:free} lists energies and entropies of the ten lowest $m=0$ levels at $\Rc=6\au$ and $F=0$, which are the starting points of the Stark maps below. At this radius the CHA shows the accidental degeneracy of the $3p$ and $4f$ levels (levels 5 and 6, $E=0.4215\au$), as predicted for $\Rc=6\au$~\cite{b146,b147,b148}.

\begin{table}[t]
\centering
\caption{Field-free energies $E$ (a.u.) and Shannon entropies of the ten lowest $m=0$ levels at $\Rc=6\au$. Levels are ordered by energy; labels are hydrogenic ($nl$ with $n=n_r+l+1$).}
\label{tab:free}
\small
\setlength{\tabcolsep}{4pt}
\begin{tabular}{c|ccccc|ccccc}
\toprule
 & \multicolumn{5}{c|}{CHA} & \multicolumn{5}{c}{PISB}\\
level & state & $E$ & $\Sr$ & $\Sp$ & $\ST$ & state & $E$ & $\Sr$ & $\Sp$ & $\ST$\\
\midrule
1 & 1s & -0.4992 & 4.105 & 2.451 & 6.557 & 1s & 0.1371 & 6.051 & 0.567 & 6.618 \\
2 & 2p & -0.0556 & 5.643 & 1.368 & 7.012 & 2p & 0.2804 & 5.896 & 1.135 & 7.030 \\
3 & 2s & 0.0127 & 6.052 & 2.155 & 8.207 & 3d & 0.4614 & 5.928 & 1.721 & 7.649 \\
4 & 3d & 0.1803 & 5.868 & 1.766 & 7.634 & 2s & 0.5483 & 5.837 & 2.356 & 8.193 \\
5 & 3p & 0.4215 & 5.600 & 2.538 & 8.138 & 4f & 0.6782 & 5.937 & 2.182 & 8.119 \\
6 & 4f & 0.4215 & 5.920 & 2.193 & 8.113 & 3p & 0.8289 & 5.631 & 2.446 & 8.077 \\
7 & 3s & 0.6318 & 5.913 & 3.567 & 9.480 & 5g & 0.9299 & 5.930 & 2.558 & 8.489 \\
8 & 5g & 0.6875 & 5.927 & 2.560 & 8.487 & 4d & 1.1489 & 5.689 & 2.786 & 8.475 \\
9 & 4d & 0.8020 & 5.663 & 2.820 & 8.483 & 6h & 1.2157 & 5.918 & 2.879 & 8.798 \\
10 & 6h & 0.9830 & 5.920 & 2.878 & 8.797 & 3s & 1.2337 & 5.761 & 3.315 & 9.076 \\
\bottomrule
\end{tabular}
\end{table}

\subsection{Static electric field: Stark maps of the energy and of the entropies}
\label{sec:stark}
\begin{figure}[p]
\centering
\includegraphics[width=0.92\textwidth]{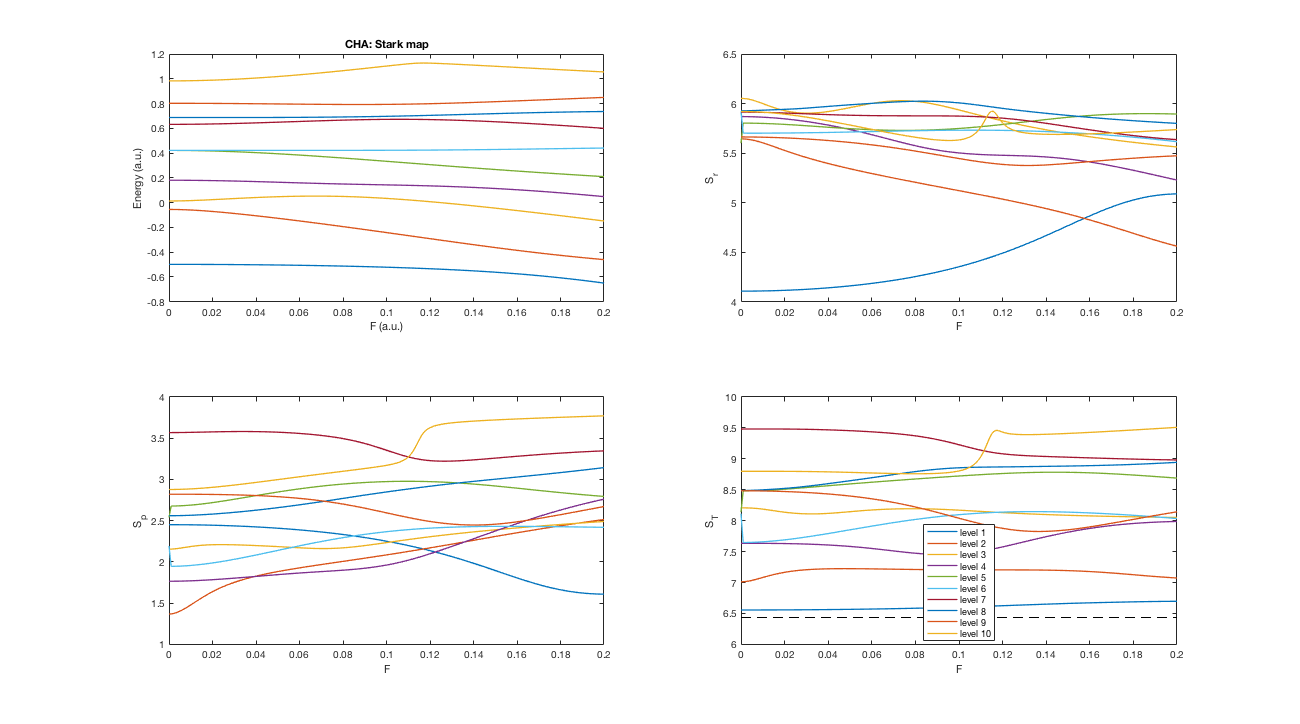}
\caption{Static-field maps of the confined hydrogen atom ($\Rc=6\au$, $m=0$): (a) energy, (b) $\Sr$, (c) $\Sp$, (d) $\ST$ of the ten lowest levels versus field strength $F$. Level numbers refer to the ordering at $F=0$ (Table~\ref{tab:free}). Dotted line in (d): BBM bound.}
\label{fig:starkCHA}
\end{figure}

\begin{figure}[p]
\centering
\includegraphics[width=0.92\textwidth]{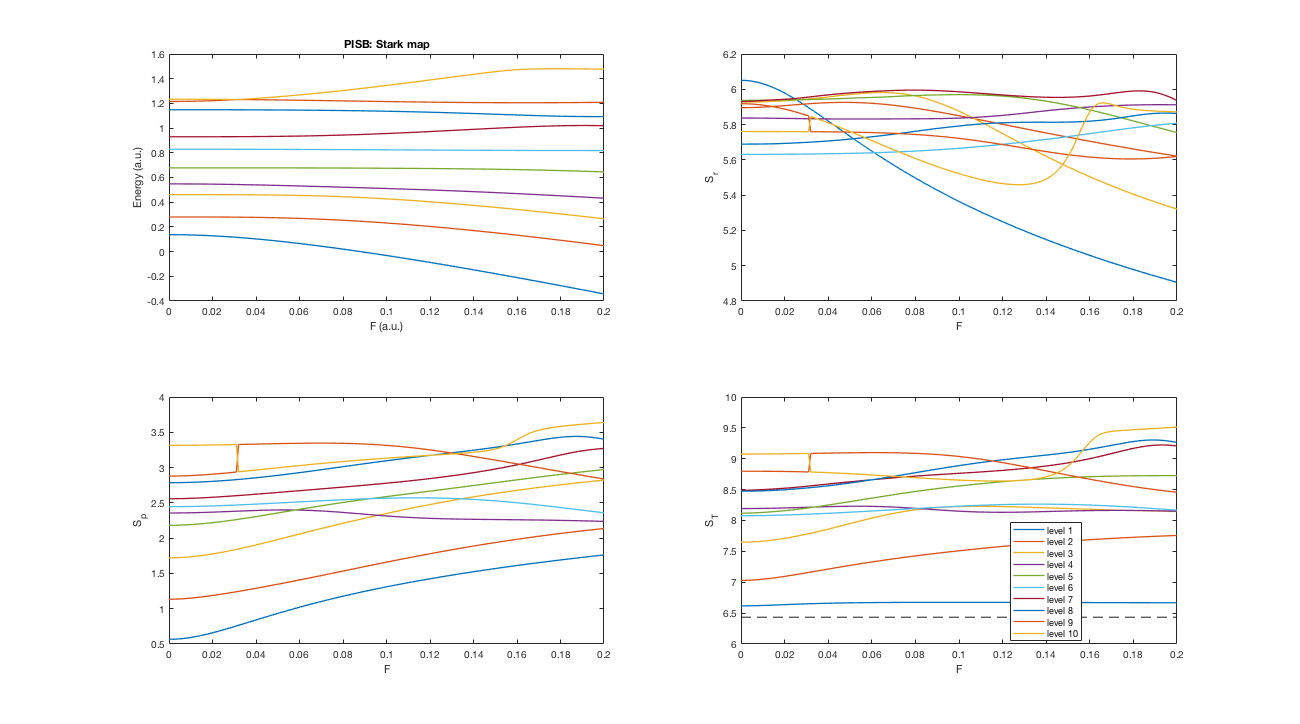}
\caption{As Fig.~\ref{fig:starkCHA} for the particle in a spherical box ($\Rc=6\au$, $m=0$).}
\label{fig:starkPISB}
\end{figure}

Figures~\ref{fig:starkCHA} and \ref{fig:starkPISB} show, for the CHA and the PISB at $\Rc=6\au$, the energies and the entropies $\Sr$, $\Sp$ and $\ST$ of the ten lowest levels as functions of the field strength $F$ up to $0.2\au$. Table~\ref{tab:field} gives the values at $F=0.2\au$ and the field-induced change of $\ST$.


\begin{table}[t]
\centering
\caption{Energies and entropies at $F=0.2\au$ ($\Rc=6\au$), and field-induced change $\delta\ST=\ST(0.2)-\ST(0)$. Level numbers are those of Table~\ref{tab:free}; at this field strength the levels have undergone several crossings and the $nl$ labels are no longer meaningful. Values for the upper levels depend on the truncation of the angular basis (see text).}
\label{tab:field}
\small
\setlength{\tabcolsep}{4pt}
\begin{tabular}{c|ccccc|ccccc}
\toprule
 & \multicolumn{5}{c|}{CHA} & \multicolumn{5}{c}{PISB}\\
level & $E$ & $\Sr$ & $\Sp$ & $\ST$ & $\delta\ST$ & $E$ & $\Sr$ & $\Sp$ & $\ST$ & $\delta\ST$\\
\midrule
1 & -0.651 & 5.088 & 1.609 & 6.697 & +0.140 & -0.343 & 4.906 & 1.762 & 6.668 & +0.050 \\
2 & -0.461 & 4.559 & 2.514 & 7.073 & +0.061 & 0.048 & 5.621 & 2.136 & 7.758 & +0.728 \\
3 & -0.149 & 5.561 & 2.490 & 8.052 & -0.155 & 0.266 & 5.321 & 2.823 & 8.144 & +0.496 \\
4 & 0.048 & 5.227 & 2.760 & 7.987 & +0.353 & 0.432 & 5.912 & 2.238 & 8.150 & -0.044 \\
5 & 0.209 & 5.894 & 2.794 & 8.688 & +0.549 & 0.645 & 5.755 & 2.972 & 8.726 & +0.607 \\
6 & 0.440 & 5.615 & 2.419 & 8.034 & -0.080 & 0.818 & 5.808 & 2.360 & 8.167 & +0.090 \\
7 & 0.600 & 5.635 & 3.346 & 8.981 & -0.499 & 1.019 & 5.936 & 3.271 & 9.208 & +0.719 \\
8 & 0.736 & 5.800 & 3.143 & 8.942 & +0.456 & 1.093 & 5.862 & 3.402 & 9.264 & +0.789 \\
9 & 0.850 & 5.472 & 2.671 & 8.143 & -0.340 & 1.209 & 5.619 & 2.839 & 8.458 & -0.340 \\
10 & 1.055 & 5.737 & 3.770 & 9.507 & +0.710 & 1.478 & 5.872 & 3.640 & 9.512 & +0.436 \\
\bottomrule
\end{tabular}
\end{table}

\paragraph{Energies and polarizability.}
In weak fields the ground-state energy decreases quadratically, $E\simeq E_0-\tfrac12\alpha F^2$. A fit to the data for $F\le0.01\au$ gives $\alpha=4.06\au$ for the CHA, about $10\%$ below the free-atom value $9/2$, and $\alpha=46.8\au$ for the PISB, an order of magnitude larger. The difference reflects the binding by the Coulomb potential: the box offers no restoring force other than the wall, and the polarizability of a box scales as $\Rc^4$. By $F=0.2\au$ the energy of level 1 has dropped by $0.151\au$ in the CHA and by $0.480\au$ in the PISB (Table~\ref{tab:field}). Level 2 of the CHA, the $2p$ state, which lies only $0.068\au$ below the $2s$ level, is far more polarizable than the ground state and moves from $-0.056\au$ to $-0.461\au$.

\paragraph{Ground-state entropy: opposite responses of $\Sr$ and $\Sp$ in the two systems.}
The most striking difference is the sign of the response of the position and momentum entropies of the ground state. In the CHA the field \emph{delocalizes} the electron in position space, $\Sr:4.105\to5.088$ ($+0.983$), and narrows its momentum distribution, $\Sp:2.451\to1.609$ ($-0.842$). In the PISB the signs are reversed: $\Sr:6.051\to4.906$ ($-1.145$) and $\Sp:0.567\to1.762$ ($+1.195$). A consistent picture is that in the box the field pushes the electron against the low-potential wall, so that the density compresses (smaller $\Sr$) and the momentum distribution broadens (larger $\Sp$), whereas in the atom the Coulomb-bound density is deformed and spread along the field. In both systems the changes of $\Sr$ and $\Sp$ nearly cancel in the sum, so that $\ST$ of the ground state changes only by $+0.140$ (CHA) and $+0.050$ (PISB), i.e.\ by $2.1\%$ and $0.8\%$. Thus $\ST$ is much less sensitive to the field than either of its parts, a point that should be kept in mind when the sum entropy alone is used as an indicator.

\paragraph{Excited levels and weak-field mixing.}
For excited levels the response of $\ST$ can be much larger. In the CHA the $\ST$ of level 2 rises quickly in weak fields, $7.012$ at $F=0$, $7.207$ at $F=0.03\au$ and $7.224$ at $F=0.05\au$, accompanied by an increase of $\Sp$ from $1.368$ to $1.752$ between $F=0$ and $0.03\au$; this is the effect of the strong Stark mixing of $2p$ with the nearby $2s$ and $3d$ levels. At larger $F$ it declines again to $7.073$. In the PISB, where the nearest level is $0.18\au$ away, the corresponding entropy changes more slowly and reaches $7.758$ at $F=0.2\au$, its maximum over the range, from $7.030$ at $F=0$. The CHA therefore shows non-monotonic entropy variations that follow from the interplay of the field with the narrower level spacing of the Coulomb problem, while the box levels, with their larger spacings, respond smoothly. All entropies remain above the BBM bound; the smallest value over all levels and fields is $6.557$ (CHA) and $6.618$ (PISB), both for the ground state at $F=0$, with margins $0.122$ and $0.183$.

\paragraph{Lifting of an accidental degeneracy.}
At $F=0$ the CHA levels 5 and 6 ($3p$ and $4f$) are degenerate at $E=0.42151\au$, with $\ST=8.138$ and $8.113$. These states differ by $\Delta l=2$ and are therefore not coupled by the dipole operator directly, but only in second order through intermediate $d$ and $g$ levels. Since there is no zero-field splitting, any non-zero field mixes them completely: already at $F=10^{-3}\au$ the entropies of the two levels have become $\ST=8.479$ and $7.647$ ($\Sr$: $5.600\to5.802$ and $5.920\to5.701$; $\Sp$: $2.538\to2.677$ and $2.193\to1.946$), although the energies are still equal to within $1.2\times10^{-5}\au$ (and differ by $3\times10^{-4}\au$ at $F=5\times10^{-3}\au$). This is an entropic exchange without a visible energy gap: the entropy discontinuity at $F\to0^+$ is a clean signature of the hybridization of an accidentally degenerate pair into field-selected combinations. In the PISB all ten levels are non-degenerate and no such discontinuity occurs (Fig.~\ref{fig:starkPISB}).

\paragraph{Avoided crossings and their entropic signature.}
Table~\ref{tab:crossings} lists the avoided crossings among adjacent levels that are resolved in the maps (defined by a minimum of the adjacent-level gap that is below $70\%$ of its value at both ends of the range), together with the change of $\ST$ of the two levels across the crossing, evaluated over a window of $\pm0.01\au$. Three features stand out.

\begin{table}[t]
\centering
\caption{Avoided crossings between adjacent levels at $\Rc=6\au$: crossing field $F_c$, minimum gap $\Delta_{\min}$, zero-field gap, difference $\Delta l$ of the angular momenta of the weak-field labels, and change of $\ST$ of the lower and upper level across $F_c\pm0.01\au$.}
\label{tab:crossings}
\small
\begin{tabular}{lcccccc}
\toprule
System & levels (states) & $F_c$ & $\Delta_{\min}$ & gap$(F{=}0)$ & $\Delta l$ & $\delta\ST$ (lower / upper)\\
\midrule
CHA  & 3--4 ($2s/3d$) & 0.079 & 0.1005 & 0.168 & 2 & $-0.00$ / $-0.06$\\
CHA  & 7--8 ($3s/5g$) & 0.093 & 0.0239 & 0.056 & 4 & $-0.16$ / $+0.05$\\
PISB & 9--10 ($6h/3s$) & 0.031 & 0.00055 & 0.018 & 5 & $+0.30$ / $-0.31$\\
\bottomrule
\end{tabular}
\end{table}

First, the width of the gap decreases with the angular-momentum difference of the states that must be connected by the field. The $2s/3d$ pair ($\Delta l=2$) is only moderately repelled, and the entropies of the two levels barely change across $F_c$: the levels approach each other without any substantial entropic exchange. The $3s/5g$ pair of the CHA ($\Delta l=4$) has a minimum gap of $0.024\au$, less than half its zero-field value, and exchanges entropy gradually: the $\ST$ of level 7 falls from $9.44$ at $F=0.06\au$ to $9.08$ at $0.12\au$ while that of level 8 rises from $8.69$ to $8.87$, with $\Sp$ carrying most of the change ($3.56\to3.23$ and $2.69\to2.92$). Second, the narrowest avoided crossing is found in the PISB between levels 9 and 10 ($6h$ and $3s$, $\Delta l=5$): at $F_c=0.031\au$ the gap is only $5.5\times10^{-4}\au$, more than thirty times smaller than at $F=0$, and across it the two levels exchange essentially all of their entropy difference, $\delta\ST=+0.30$ and $-0.31$ against a diabatic difference of $0.28$, with $\Sp$ changing by $+0.42$ and $-0.35$ and $\Sr$ by $-0.13$ and $+0.04$. This is a nearly diabatic, abrupt exchange of the kind that makes the entropy a sensitive indicator of avoided crossings~\cite{b49}. Third, the CHA level 10 shows a sharp step in $\Sp$ and $\ST$ near $F\simeq0.112$--$0.117\au$ ($\ST$ from $8.80$ at $F=0.10\au$ to $9.40$ at $0.125\au$; $\Sp$ from $3.17$ to $3.67$), together with a turning point in its energy at $0.117\au$; we attribute this to an avoided crossing with a level just above the ten shown, whose partner is therefore not displayed, and since it involves the highest level it is also sensitive to the truncation of the basis.

Other approaches of adjacent levels (for instance levels 1 and 2 of the CHA, or $2p/3d$ of the PISB) show a gap that varies smoothly without a pronounced minimum and no sharp entropy signature; their $\ST$ varies smoothly as well. We conclude that a clear entropic exchange requires both a small gap and a coupling that is weak enough for the crossing to be nearly diabatic, which is the content of Eq.~(\ref{eq:twolevel}).

\subsection{Time-dependent field: laser-induced avoided crossings}
\label{sec:laser}
\begin{figure}[t]
\centering
\includegraphics[width=\textwidth]{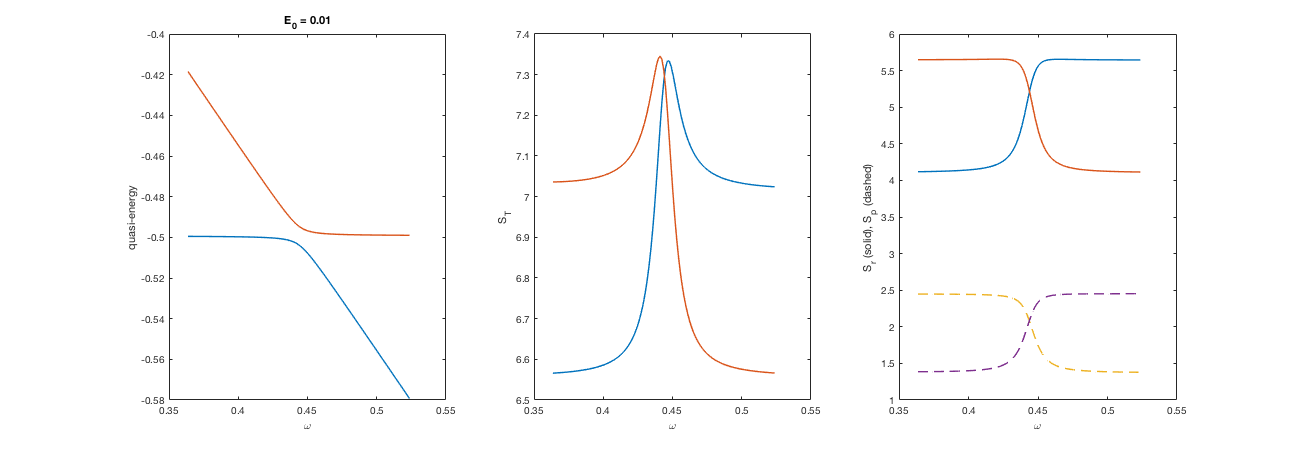}
\caption{Laser-dressed confined hydrogen atom ($\Rc=6\au$) at amplitude $E_0=0.01\au$ near the one-photon $1s$--$2p$ resonance: (a) quasi-energies, (b) total Shannon entropy $\ST$, (c) $\Sr$ (solid) and $\Sp$ (dashed) of the two Floquet branches versus $\omega$. Entropies are those of the cycle-averaged density, Eq.~(\ref{eq:cycleavg}); dotted line in (b): BBM bound.}
\label{fig:laser1}
\end{figure}

\begin{figure}[t]
\centering
\includegraphics[width=\textwidth]{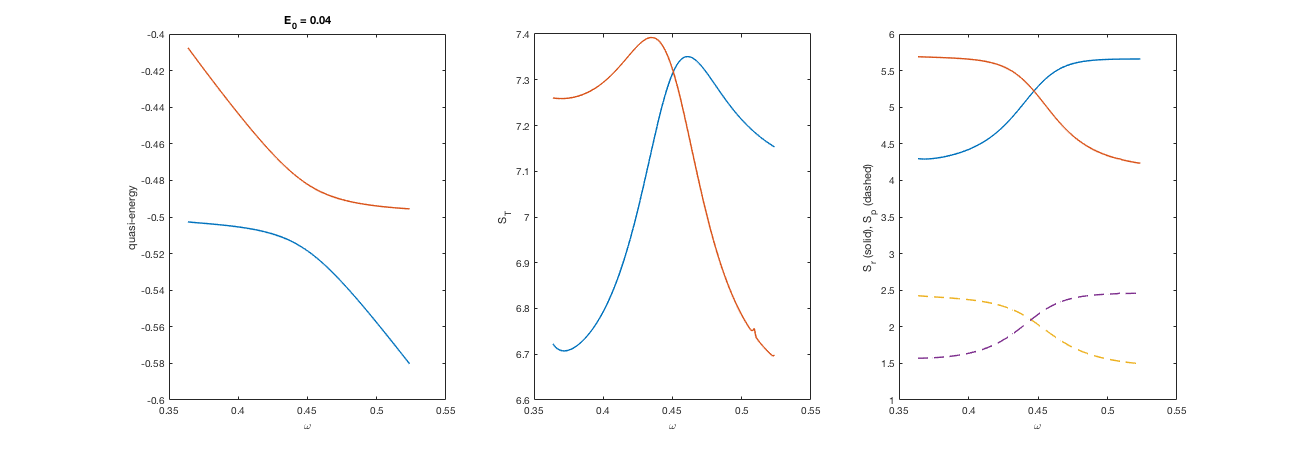}
\caption{As Fig.~\ref{fig:laser1} for $E_0=0.04\au$.}
\label{fig:laser2}
\end{figure}

For the laser-driven atom we use the Floquet formalism of Sec.~\ref{sec:methods}. The levels that take part in the interaction are selected by the dipole selection rule $\Delta l=\pm1$: the $1s$ ground state couples by one photon only to $p$ states, and the lowest of these is $2p$ ($E_{2p}-E_{1s}=0.4437\au$ at $\Rc=6\au$). Figures~\ref{fig:laser1} and \ref{fig:laser2} show the quasi-energies and the cycle-averaged entropies of the two Floquet branches that cross at this resonance, as a function of $\omega$, for $E_0=0.01$ and $0.04\au$; Table~\ref{tab:laser} summarizes them. For $\omega$ well below resonance branch 1 is the $1s$-like state (zero-photon component dominant) and branch 2 the $2p$-like state dressed by one photon (quasi-energy $E_{2p}-\omega$); above resonance their roles are interchanged.


\begin{table}[t]
\centering
\caption{Properties of the laser-induced avoided crossing between the $1s$-like and $2p$-like Floquet branches of the CHA at $\Rc=6\au$. Window of frequencies: $0.364\le\omega\le0.524\au$. The 10--90\% widths are for branch 1.}
\label{tab:laser}
\small
\begin{tabular}{lcc}
\toprule
 & $E_0=0.01$ & $E_0=0.04$\\
\midrule
$E_0$ (a.u.) & 0.01 & 0.04 \\
$\omega_c$ (a.u.) & 0.4437 & 0.4490 \\
$\Delta_{\min}$ (a.u.) & 0.00922 & 0.03636 \\
$\Delta_{\min}/E_0$ & 0.922 & 0.909 \\
$S_T$ far below resonance (branch 1 / 2) & 6.566 / 7.036 & 6.723 / 7.260 \\
$S_T$ far above resonance (branch 1 / 2) & 7.024 / 6.566 & 7.153 / 6.698 \\
$S_T$ at $\omega_c$ (branch 1 / 2) & 7.283 / 7.312 & 7.306 / 7.332 \\
maximum $S_T$ in window & 7.346 & 7.392 \\
10--90\% width of $S_r$ exchange (a.u.) & 0.023 & 0.064 \\
10--90\% width of $S_p$ exchange (a.u.) & 0.025 & 0.080 \\
\bottomrule
\end{tabular}
\end{table}

\paragraph{Gap and resonance.}
For $E_0=0.01\au$ the minimum separation of the quasi-energies is $\Delta_{\min}=9.22\times10^{-3}\au$ at $\omega_c=0.4437\au$, which coincides with the field-free transition frequency. The ratio $\Delta_{\min}/E_0=0.922$ equals the $1s$--$2p$ dipole matrix element $d_{12}=\langle1s|z|2p\rangle=0.9225\au$ (Eq.~\ref{eq:dipole}), as expected for a Rabi splitting $\Delta=E_0d_{12}$ in the rotating-wave picture. At $E_0=0.04\au$ the gap is $0.0364\au$, still close to the linear prediction ($\Delta_{\min}/E_0=0.909$, i.e.\ $1.5\%$ below $d_{12}$), but the crossing has moved to $\omega_c\simeq0.449\au$, a blue shift of about $0.005\au$ ($1.2\%$). This is the AC Stark effect. Restricting the Floquet space step by step shows where the shift comes from: about $0.001\au$ is the counter-rotating (Bloch--Siegert) shift of the two-level system, another $0.001\au$ comes from the remaining $s$ and $p$ levels, and the dominant part, about $0.003\au$, from the coupling to $d$ levels, which shifts the $2p$ quasi-energy relative to that of $1s$ (the $1s$ level couples only to $p$ states).

\paragraph{Entropic exchange.}
Far below the resonance the two branches carry the entropies of the parent states, $(\Sr,\Sp,\ST)=(4.118,\,2.448,\,6.566)$ for branch 1 ($1s$-like) and $(5.649,\,1.387,\,7.036)$ for branch 2 ($2p$-like) at $E_0=0.01\au$, values that are close to the field-free ones of Table~\ref{tab:free} ($1s$: $4.105,\,2.451,\,6.557$; $2p$: $5.643,\,1.368,\,7.012$). Across the avoided crossing the branches interchange these values: for branch 1 $\Sr$ goes from $4.118$ to $5.646$ and $\Sp$ from $2.448$ to $1.379$, i.e.\ essentially the full difference between the two bare states, so the exchange of the position and momentum entropies is nearly complete. For $\Sr$ and $\Sp$ the exchange is monotonic and takes place over $\Delta\omega_{10\text{--}90}=0.023$ and $0.025\au$. The two-level model of Eq.~(\ref{eq:twolevel}), with the diabatic energy difference changing at unit rate with $\omega$, predicts $8\Delta/3=0.025\au$ for $\Delta=9.22\times10^{-3}\au$, in agreement with the numerical widths to within $8\%$. The width of the entropic exchange is thus set by the Rabi gap.

\paragraph{Non-additivity of $\ST$ at resonance.}
The total entropy does not follow the monotonic path of its parts. Near $\omega_c$ the sum $\ST$ of \emph{both} branches rises above the values of both parent states: at $E_0=0.01\au$ it is $7.283$ and $7.312$ at $\omega_c$ and reaches a maximum of $7.346$, compared with $6.566$ and $7.036$ for the parents. The same occurs in $\Sr$ and $\Sp$ separately (at $\omega_c$, $\Sr\simeq5.20$--$5.23$ and $\Sp\simeq2.08$ against the means $4.88$ and $1.91$ of the parents). At resonance each dressed state is an equal-weight superposition of $1s$ and $2p$, and the interference (cross) term in the density discussed after Eq.~(\ref{eq:twolevel}) produces an $sp$ hybrid that is more spread out in both position and momentum space than either parent. This shows that the entropy of a coherent superposition is not a weighted average of the entropies of its components; in this sense the entropic content of the dressed state is a more sensitive probe of the coherence induced by the laser than the quasi-energies alone.

\paragraph{Influence of the laser amplitude.}
Raising the amplitude by a factor of four (Fig.~\ref{fig:laser2}) has three effects. (i) The gap grows by a factor $3.9$ (Table~\ref{tab:laser}) and the exchange widens correspondingly: the 10--90\% widths of $\Sr$ and $\Sp$ increase from $0.023$ and $0.025\au$ to $0.064$ and $0.080\au$. The exchange is more gradual and the abrupt change observed at weak amplitude is replaced by a smooth variation that extends over a large part of the frequency window. (ii) The entropies of the branches far from resonance are shifted by the stronger dressing: for $\omega=0.364\au$ the $1s$-like branch has $\ST=6.723$ instead of $6.566$, with $\Sr=4.298$ ($+0.18$) while $\Sp$ is almost unchanged ($2.425$), and the $2p$-like branch has $\ST=7.260$ instead of $7.036$. The field thus admixes $p$ and $d$ components even at a detuning of $0.08\au$, and the entropies of the dressed states are no longer those of the bare states. (iii) The overshoot of $\ST$ at the crossing grows, to a maximum of $7.392$, and the curves of the two branches develop a pronounced asymmetry, which we attribute to the AC Stark shifts of the $1s$ and $2p$ levels making the diabatic energy difference depend on $\omega$ in a different manner on the two sides of the resonance. We note that the width ratio $\Delta\omega_{10\text{--}90}/\Delta_{\min}$ is $2.5$ and $2.7$ for $\Sr$ and $\Sp$ at $E_0=0.01\au$ but only $1.8$ and $2.2$ at $E_0=0.04\au$, so that the simple linear two-level picture is quantitatively valid only for weak amplitudes. Hence the AC Stark effect, which becomes important at higher intensity, modifies not only the position and width of the avoided crossing but also the way in which $\Sr$ and $\Sp$ are exchanged.

\section{Static versus time-dependent fields}
\label{sec:compare}
The results of Secs.~\ref{sec:stark} and \ref{sec:laser} allow a direct comparison of the two kinds of field. In both cases the entropic exchange takes place where two adiabatic states exchange character, and in both cases its profile follows the mixing angle of Eq.~(\ref{eq:twolevel}); the differences lie in what plays the role of the control parameter, which states are coupled, and what the entropies refer to.

\begin{table}[t]
\centering
\caption{Comparison of the entropic response to a static and a time-dependent (laser) field, as found for the CHA and PISB at $\Rc=6\au$.}
\label{tab:compare}
\small
\renewcommand{\arraystretch}{1.25}
\begin{tabular}{p{3.0cm}p{5.8cm}p{5.8cm}}
\toprule
 & Static field (Stark map) & Laser field (Floquet)\\
\midrule
Control parameter & field strength $F$ & frequency $\omega$ at fixed amplitude $E_0$\\
Objects that cross & energy levels (eigenvalues of $H_0+Fz$) & quasi-energies of Floquet branches (state plus $m$ photons)\\
Coupling & field-induced mixing of $l$ and $l\pm1$; levels with $\Delta l>1$ connected only in higher order, giving very small gaps & dipole coupling $E_0d_{12}/2$ between photon-dressed states; one-photon resonances restricted by $\Delta l=\pm1$\\
Gap & depends on the pair; from $0.10\au$ ($\Delta l=2$) down to $5.5\times10^{-4}\au$ ($\Delta l=5$) & Rabi splitting $E_0d_{12}$: $9.2\times10^{-3}\au$ at $E_0=0.01\au$, $3.6\times10^{-2}\au$ at $0.04\au$\\
Entropy of a state & of the stationary eigenstate & of the cycle-averaged density of the Floquet state\\
Ground-state response & $\Sr$ and $\Sp$ change by about $\pm1$, $\ST$ by $0.05$--$0.14$ at $F=0.2\au$ & at resonance both branches are $1s$--$2p$ hybrids with $\ST\simeq7.3$\\
Broadening & set by the gap; narrow for large $\Delta l$ & set by the Rabi gap, grows with $E_0$ together with the AC Stark shift\\
Special features & discontinuous entropy change at $F\to0^+$ for accidentally degenerate levels & overshoot of $\ST$ at resonance; shift of $\omega_c$ with intensity\\
\bottomrule
\end{tabular}
\end{table}

Three points deserve emphasis. First, a static field has a single control parameter and the Stark map contains many crossings, each with its own gap that is controlled by the angular-momentum difference of the levels; the entropy then behaves as a sensitive marker of the narrowest, nearly diabatic crossings and of the lifting of degeneracies. In the laser case the selection rules pick out one resonant pair, and the gap is controlled by the amplitude, so that the strength of the entropic exchange can be tuned at will; the AC Stark shift moves the position of the crossing and, at higher amplitude, smoothes the exchange. Second, the ground-state entropies of the confined atom and of the box respond to a static field with opposite signs in $\Sr$ and $\Sp$, while in the laser case what matters is the coherence between the two resonant states. Third, the sum entropy is much less sensitive than $\Sr$ and $\Sp$ separately for a state that is merely polarized by the field, but it is the quantity that registers coherent superposition by its non-additive overshoot.

Comparing the two confinement models, the difference between the CHA and the PISB is most pronounced for properties that depend on polarizability (a factor $\simeq11$ at $\Rc=6\au$) and for the field-free scale dependence ($\ST$ is $\Rc$-independent for the box but not for the atom). The box is therefore a useful limiting case in which the entropies are determined solely by the nodal structure and the field, while the atom shows the additional competition between Coulomb binding and confinement, including accidental degeneracies at special radii.

\section{Conclusions and outlook}
\label{sec:concl}
We have reviewed and extended the study of external-field effects on the information measures of confined atoms. For the field-free systems the total Shannon entropy of the particle in a spherical box is independent of the box radius, while that of the confined hydrogen atom depends non-monotonically on it and approaches the box values for strong confinement. In a static electric field the ground-state polarizability, and with it the entropic response, differs by an order of magnitude between the atom and the box, and the signs of the changes of $\Sr$ and $\Sp$ are opposite, although the sum $\ST$ changes little. Accidentally degenerate levels exchange entropy discontinuously as soon as a field is applied, and avoided crossings between levels of very different angular momentum are extremely narrow and give an almost complete exchange of the entropies. For a monochromatic laser field the Floquet formalism shows that the one-photon resonance produces an avoided crossing with a gap equal to the Rabi splitting, across which the position and momentum entropies are exchanged over a frequency range proportional to the gap, and where the total entropy of the dressed states exceeds that of the bare states. At larger intensity the AC Stark effect shifts the crossing and broadens the exchange. Static and time-dependent fields therefore produce related but distinguishable entropic signatures, which establishes a framework for telling their effects apart. Throughout, the BBM bound is respected, and the ground state of the confined atom stays closest to it.

Several extensions are suggested by this work. The soft-wall (fullerene) confinement and Neumann-type boundary conditions should be studied with the same tools to see how penetration into the wall changes the exchange; further information measures (Onicescu energy, Kullback--Leibler distance, R\'enyi, Tsallis and LMC measures) can be tracked across the same crossings; the field-dependent entropies should be extended to multi-electron systems and to combined electric and magnetic fields; and a genuinely time-dependent Shannon entropy for pulses, rather than the cycle average used here, would connect the present results to dynamical processes. Because the entropy of a field-dressed state reports on its coherence and delocalization, such studies may be useful for the design and control of superposition states in quantum information science.

\bigskip
\noindent\textbf{Data and reproducibility.} The figures of Secs.~\ref{sec:free}--\ref{sec:laser} are drawn from the numerical results described in Sec.~\ref{sec:methods}; the values quoted in the text and in the tables are read from the same data.

\end{document}